\documentclass[aps,prd,twocolumn,groupedaddress,showpacs]{revtex4}

\usepackage{graphicx}
\usepackage{dcolumn}
\begin{document}

\title{\bf \boldmath On the analysis of the $\pi\to e\nu\gamma$ experimental 
       data}
\author{A.A.~Poblaguev}
\email{poblaguev@bnl.gov}
\affiliation{Physics Department, Yale University, New Haven, CT 06511}
\affiliation{Institute for Nuclear Research of the Russian Academy of Sciences, 117312 Moscow, Russia}

\date{July 11, 2003}

\begin{abstract}
 The most general amplitude for
 the radiative pion decay $\pi\to e\nu\gamma$ including terms beyond V-A 
 theory is considered. 
 The experimental constraints on the decay amplitude components are
 discussed. A model independent presentation of the results of
 high statistics and high resolution experiments is suggested.
\end{abstract}


\maketitle

\begin{figure}
\begin{center}
\mbox{\includegraphics[width=0.45\textwidth]{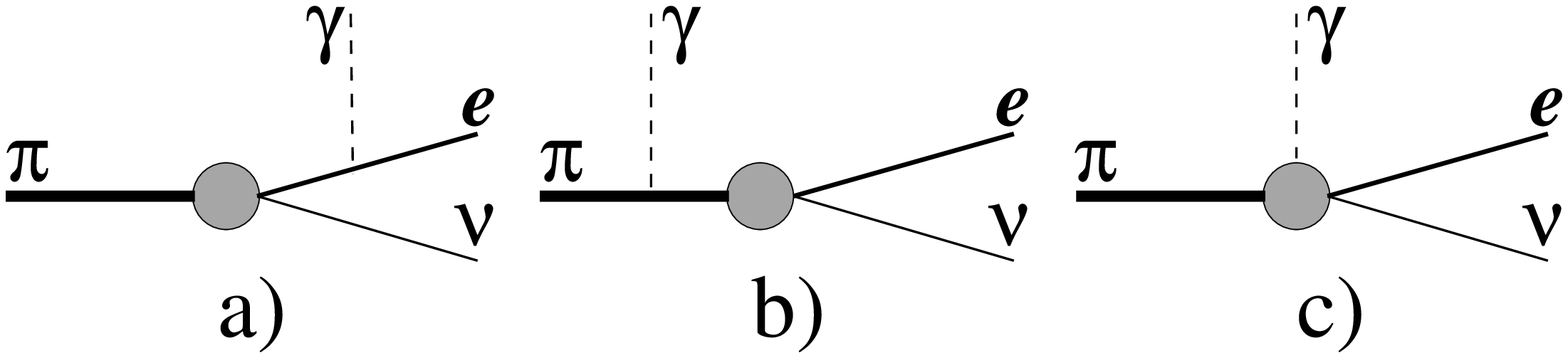}}
\end{center}
\caption{The $\pi\to e\nu\gamma$ diagrams}
\label{graph}
\end{figure}

Due to the strong helicity suppression of the $\pi\to e\nu$ decay, the
corresponding radiative decay
\begin{equation}
   \pi \to e \nu \gamma
\end{equation}
is regarded as a valuable source of pion structure information and a
sensitive test anomalous interactions
\cite{bardin,review}.  
Low's theorem \cite{low} allows the part of the $\pi\to
e\nu\gamma$ amplitude depending on the photon energy $\omega$ as
$\omega^{-1}$ and $\omega^0$ to be  unambiguously defined by the pion decay
constant $f_\pi=131~\mathrm{MeV}$.  This contribution to the decay
amplitude, called inner bremsstrahlung ($IB$), is associated with
the tree diagrams shown in Figs. \ref{graph}a) and \ref{graph}b). 
Generally, one can construct the
remaining, structure dependent ($SD$) part of the amplitude
(Fig. \ref{graph}c)) using the  
photon tensors ${\cal
F}^{\mu\nu}_{\pm}=k^\mu\epsilon^\nu-k^\nu\epsilon^\mu\pm i
e^{\mu\nu\alpha\beta}k^{\alpha}\epsilon^\beta$ where $\epsilon^\mu$ is the
photon polarization vector and  $\pm$ specify the
photon helicities, the lepton currents
$L         =\bar{u}_\nu(1+\gamma^5)                u_e$,  
$L^\mu     =\bar{u}_\nu(1+\gamma^5)\gamma^\mu      u_e$,   
$L^{\mu\nu}=\bar{u}_\nu(1+\gamma^5)\sigma^{\mu\nu} u_e$, and three
available 4-momenta
$k^\mu$, $q^\mu=l^\mu+t^\mu$, and $r^\mu=l^\mu-t^\mu$, where $k^\mu$, $l^\mu$, and $t^\mu$ are 4-momenta of photon, electron, and neutrino, respectively.
 Using the Dirac
equation and properties of $\gamma$-matrices the full
list of such amplitudes:
${\cal F}^{\mu\nu}_\pm a^\mu b^\nu L$, 
${\cal F}^{\mu\nu}_\pm a^\mu b^\nu L^\rho c^\rho$, 
${\cal F}^{\mu\nu}_\pm a^\mu b^\nu L^{\rho\sigma} c^\rho d^\sigma$, 
${\cal F}^{\mu\nu}_\pm a^\mu L^\nu$, 
${\cal F}^{\mu\nu}_\pm a^\mu b^\rho L^{\nu\rho}$, and 
${\cal F}^{\mu\nu}_\pm L^{\mu\nu}$,  
may be reduced to
four amplitudes:
\begin{eqnarray}
   m_\pi {\cal V_{\pm}} & = & 
      {\cal F}^{\mu\nu}_\pm q^\mu L^\nu \label{Vpm} \\
         {\cal T}       & = & 
      {\cal F}^{\mu\nu}_-L^{\mu\nu}/2 = k^\mu\epsilon^\nu L^{\mu\nu}
\label{T} \\
   m_\pi^2 {\cal S}     & = &
      {\cal F}^{\mu\nu}_+q^\mu (q^\rho L^{\rho\nu}+m_eL^\nu) 
      = {\cal F}^{\mu\nu}_+q^\mu r^\nu L
\label{S} 
\end{eqnarray}
where $m_e$ and $m_\pi$ are the electron and pion masses, respectively.

With these ingredients, the full $\pi\to e\nu\gamma$ amplitude may be presented as :\\
\begin{eqnarray}
& & {\cal M}_{\pi\to e\nu\gamma} =\frac{-ieG_FV_{ud}}{\sqrt{2}} \biggl\{ 
  \frac{f_\pi}{m_\pi}\frac{\sqrt{r_e}}{x^2\lambda}
  \Bigl({\cal S}+{\cal T}+ \sqrt{r_e}{\cal V_-} \Bigr) +  \nonumber \\
 &  & 
 ~~~~~~~~\frac{F_V+F_A}{2}{\cal V_+} + \frac{F_A-F_V}{2}{\cal V_-} + F_T {\cal
  T} + F_S {\cal S} \biggr\}
\label{amplitude}
\end{eqnarray}
The first term in the sum is the $IB$ amplitude. The structure dependent
contributions are parameterized by form factors $F_V$, $F_A$, $F_T$, and
$F_S$. In Eq. (\ref{amplitude}), the vector and axial-vector form factors,
$F_V$ and $F_A$, are defined in accordance with the Particle Data Group 
(PDG)\cite{PDG} and the tensor form factor $F_T$ uses the definition of
Ref. \cite{tensor}.
In the most general case, form factors 
are functions of the kinematic variables  $x=2kq/m_\pi^2$ and 
$\lambda=kl/kq$. $\lambda$ is expressed in terms of photon, 
$x=2E_\gamma/m_\pi$, 
and electron, $y=2E_e/m_\pi$, center of mass energies as 
$\lambda=(x+y-1-r_e)/x$, 
where $r_e=m_e^2/m_\pi^2$. Neglecting the electron  mass $m_e$, one can
relate $\lambda$ to the $e$-$\gamma$ opening angle in a simple form,
$\lambda =y\sin^2(\theta_{e\gamma}^{(\pi)}/2) = \sin^2(\theta_{e\gamma}^{(e\nu)}/2)$, 
where $\theta_{e\gamma}^{(\pi)}$ and $\theta_{e\gamma}^{(e\nu)}$ are
angles between $e$ and $\gamma$ momenta in the pion and lepton pair
centers of mass,  respectively.
In this paper we will assume that the form factors do not depend on
$\lambda$. Depending on the possible origin of the amplitudes, {\em e.g.} 
${\cal F}^{\mu\nu}_-q^\mu q^\rho L^{\nu\rho}\propto(1-x){\cal T}$,
${\cal F}^{\mu\nu}_-q^\mu k^\rho L^{\nu\rho}\propto x{\cal T}$,
the form factors may have a strong dependence on $q^2$ or ,
equivalently, on $x$.

To calculate the $\pi\to e\nu\gamma$ decay rate, the following sums
over polarizations can be used: 
\begin{eqnarray}
   \rho_{\cal V_+\phantom{S} } & = & 
      x^2(1-x)\lambda^2            - r_e x^2\lambda             \\
   \rho_{\cal V_-\phantom{T} } & = & 
      x^2(1-x)(1-\lambda)^2        + r_e x^2(1-\lambda)         \\
   \rho_{\cal T\phantom{V_-} } & = & 
      x^2\lambda(1-\lambda)                                     \\
   \rho_{\cal S\phantom{V_+} } & = & 
      x^2(1-x)^2\lambda(1-\lambda)                   \nonumber  \\
      & & - r_e x^2(1-\lambda)[(1-x)(1+\lambda)-r_e]\\
   \rho_{\cal TV_-}            & = & 
      \sqrt{r_e}x^2(1-\lambda)                                  \\
   \rho_{\cal SV_+}            & = & 
      \sqrt{r_e}x^2(1-\lambda) [(1-x)\lambda - r_e]             \\
   \rho_{\cal SV_-} & = & 
      \rho_{\cal ST}=\rho_{\cal TV_+}=\rho_{\cal V_+V_-}=0
\end{eqnarray}
where  $\rho_X=\sum_{pol}|X|^2/(8m_\pi^4)$ and 
$\rho_{XY}=\sum_{pol}|XY^*|/(8m_\pi^4)$. In the limit $m_e=0$, each amplitude 
${\cal V_\pm}$, ${\cal T}$, ${\cal S}$ is characterized by definite photon and electron helicities  resulting the absence of the interference with each 
other.
In this approximation, the decay rate, normalized to the branching ratio,
 may be presented as
\begin{eqnarray}
 \frac{d^2\mathrm{BR}_{\pi\to e\nu\gamma}}{dx d\lambda}  & = & 
 1.43\cdot10^{-7}\times
\biggl\{ 
        \frac{(1-x)^2+1}{x}\frac{1-\lambda}{\lambda} \nonumber \\
       & + & |f_V+f_A|^2x^3(1-x)\lambda^2            \nonumber \\
       & + & |f_V-f_A|^2x^3(1-x)(1-\lambda)^2        \nonumber \\
       & + & 4\Big[f_T+f_S(1-x)^2\Big]x(1-\lambda)   \nonumber \\
       & + & 4\Big[f_T^2+f_S^2(1-x)^2\Big]x^3\lambda(1-\lambda)
\biggr\}
\label{rate}
\end{eqnarray}
In Eq. (\ref{rate}) we have scaled the form factors 
\begin{equation}
f_{V,A,T,S}=(m_\pi^2/2m_ef_\pi)F_{V,A,T,S}\approx146F_{V,A,T,S}
\end{equation}
 The first, $IB$, term
is free of unknown parameters. The terms proportional to $|f_V\pm
f_A|^2$ are traditionally referred as $SD^\pm$ terms. 
The term proportional to $f_T+f_S(1-x)^2$ describes the interference
of the amplitudes $\cal T$ and  $\cal S$ with $IB$.

In the Standard Model $F_T=F_S=0$. The $q^2=(1-x)m_\pi^2$ variation of the 
form factors $F_V(q^2)$ and $F_A(q^2)$ is not expected to exceed a few
percent over the 
total $\pi\to e\nu\gamma$ phase space \cite{bardin,bijnens,geng}.
The conservation of vector current (CVC) hypothesis relates
the vector form factor $F_V$ to the lifetime of the neutral pion
\begin{equation}
|F_V(0)| = 
\frac{1}{\alpha}\sqrt{\frac{2\Gamma(\pi^0\to\gamma\gamma)}{\pi m_{\pi^0}}}
=0.0259\pm0.0005
\label{FV}
\end{equation}
\cite{vaks} or equivalently  $f_V\approx3.78$.
As result, the $\pi\to e\nu\gamma$ decay amplitude contains
only one unknown parameter 
$\gamma=F_A/F_V$ which has to be determined empirically. 

The $\pi\to e\nu\gamma$ experiments \cite{cern, berkeley,lampf,psi}, which 
used the stopped pion technique, 
 measured decay only in a limited phase space region where
$SD^+$ contribution  is strongly dominant. 
By counting the $\pi\to e\nu\gamma$ decays one could determine only
$|1+\gamma|^2$ with a small correction due to $|1-\gamma|^2$.
Some extension of the investigated phase space region \cite{lampf,psi} allowed
 one to resolve the ambiguity in the sign of $\gamma$ and find a unique value:
\begin{equation}
	\gamma = 0.41\pm0.06
\label{gamma}
\end{equation}
\cite{PDG}. 
This result is derived with the assumption that the value of the vector form factor in Eq. (\ref{FV}) is valid. The choice of the sign of $\gamma$ was also confirmed by the $\pi^+\to e^+\nu e^+e^-$ measurement \cite{sindrum}.

The ISTRA experiment \cite{istra,istra_yaf}, in which pion decays were studied
in flight, investigated the
$\pi\to e\nu\gamma$ events over a much larger phase space region:
$0.3<x<1,~0.2<\lambda<1$. The measured $\pi\to e\nu\gamma$ branching ratio
$1.61\pm0.23$, in units of $10^{-7}$, was found significantly smaller than the 
expected sum $2.41\pm0.07$ of the $IB$ (1.70), $SD^+$ ($0.67\pm0.07$), and $SD^-$ (0.04) contributions. 
The fit to the experimental distribution  by the linear combination of the 
functions  $IB(x,\lambda)$ and $SD^\pm(x,\lambda)$ indicated a non-physical 
contribution of the $SD^-$, 
$\mathrm{BR}_{SD^-}=-0.58\pm0.20$, to the total branching ratio.
 This result was interpreted \cite{tensor}
as a possible indication of a tensor amplitude with $F_T=-0.0056\pm0.0017$
$(f_T=-0.82\pm0.25)$. It was shown \cite{confirm} that this hypothesis not 
only does not contradict the results of the previous experiments 
\cite{cern,berkeley,lampf,psi} but even improves their consistency.

Analysis of the new high statistics
measurement of about 70,000 $\pi\to e\nu\gamma$ decays performed by the PIBETA
collaboration
\cite{pibeta} is currently in progress. The preliminary results also
indicate a  
deficit in the observed $\pi\to e\nu\gamma$ decays \cite{frlez},
however, a 2-parameter fit for $F_A$ and $F_T$ in the phase space region 
$E_\gamma>55.6~\mathrm{MeV}$ and $E_e>20~\mathrm{MeV}$ gives a much
smaller value for $F_T=-0.0017(1)$  ($f_T=-0.25\pm0.02$) \cite{pocanic}.

The PIBETA preliminary result calls for a more accurate study of the possible 
phenomenological explanation  
of the anomaly in the $\pi\to e\nu\gamma$ branching ratio. 
In this analysis we will assume that the discrepancy in the results are 
not caused by the experimental error nor by the statistical fluctuation. 
Since the measured number of events is smaller then expected, it can not be
explained by some unknown background. Modification of the decay
amplitude is required. 
To make an estimate, one can integrate Eq. (\ref{rate}) over the ISTRA
phase space region
assuming constant form factors:
\begin{eqnarray}
& & 10^7\mathrm{BR}_{\pi\to e\nu\gamma}  =  1.70 +
                      0.327|1+\gamma|^2+0.169|1-\gamma|^2 \nonumber \\
& & ~~~~~+   0.833f_T + 0.212f_T^2  +   0.099f_S + 0.013f_S^2
\label{rateISTRA}
\end{eqnarray}
Depending on the value of $\gamma$, the total $SD^\pm$ contribution to the decay
rate can vary from $0.61$ for $\gamma=0.25$ \cite{lampf} to $0.79$ for
$\gamma=0.52$ \cite{psi}. Any assumptions about the form factors
$F_V$ and $F_A$, {\em e.g.} strong dependence on $q^2$, violation of CVC, big
imaginary parts can not significantly change this estimate, because
in Ref. \cite{berkeley} they measured the structure dependent
contribution to be $0.39\pm0.07$
in the $E_e>56~\mathrm{MeV}$, $E_\gamma>30~\mathrm{MeV}$,
$\theta^{(\pi)}_{e\gamma}>132^\circ$ region of the ISTRA total phase space.
This value establishes a lower bound of  $2.1\times10^{-7}$ for the
$\pi\to e\nu\gamma$ branching ratio if
$F_T=F_S=0$.
Disagreement with the ISTRA result $(1.61\pm0.23)\times10^{-7}$ clearly
indicates the necessity to consider $\cal T$ and $\cal S$ contributions.

\begin{table*}
\caption{\label{tabl} The comparison of form factors $f_T(x)$
and $f_S(x)$ with the available experimental data. The 
form factors are parameterized by the constant $f$. For the combined
analysis of the experiments \cite{cern,berkeley,lampf,psi}, the
estimate of $f$ and the corresponding consistency of the data $\chi^2$
are shown.
For ISTRA, $\chi^2$ indicate the agreement between the
simulated and measured values of $BR_{IB}$ and
$BR_{SD^-}$. Predictions for the systematic shift of 
$\gamma$) $\Delta\gamma$ and $f_T$, which one can obtain from the preliminary PIBETA fit
are based on the estimation of the $f$ in the simulation of the ISTRA 
fit. The predicted values of the $f_T$ may be compared with the preliminary 
experimental result $f_T\approx0.25$ \cite{pocanic}.
}
\begin{ruledtabular}
\begin{tabular}{cc|dd|dd|dd}
\multicolumn{2}{c|}{Considered form factors} & 
\multicolumn{2}{c|}{Experiments \cite{cern,berkeley,lampf,psi}} &
\multicolumn{2}{c|}{ISTRA} & 
\multicolumn{2}{c}{Predictions for the PIBETA fit}
\\
$f_T$ & $f_S$ & \multicolumn{1}{c}{$f$}            & \multicolumn{1}{c|}{$\chi^2$}
              & \multicolumn{1}{c}{$f$}            & \multicolumn{1}{c|}{$\chi^2$}
              & \multicolumn{1}{c}{$\Delta\gamma$} & \multicolumn{1}{c}{$f_T$}
\\ \hline
 $0$        & $ 0$     & \multicolumn{1}{c}{---} &  5.66   
                       & \multicolumn{1}{c}{---} & 18.98   
                       & 0.0                     &  0.00 
\\ 
 $f$        & $ 0$     & -1.4^{+0.9}_{-0.7}      &  2.97   
                       & -1.7\pm0.5              &  0.50   
                       &  0.0                    & -1.68^{+0.47}_{-0.51} 
\\
 $fx$       & $ 0$     & -1.6^{+1.0}_{-0.6}      &  2.92   
                       & -2.8^{+0.8}_{-1.3}      &  0.05   
                       & -0.09^{+0.07}_{-0.02}   & -2.52^{+0.76}_{-1.12} 
\\
 $f(1-x)$   & $ 0$     & -5.5^{+3.7}_{-3.0}      &  2.89   
                       & -4.1^{+1.2}_{-1.3}      &  4.40   
                       &  0.01                   & -0.51^{+0.16}_{-0.19} 
\\
 $f(1-x)^2$ & $ 0$     &-18.4^{+12.3}_{-6.8}     &  3.10  
                       & -6.6^{+2.3}_{-2.6}      &  9.45   
                       &  0.0                    & -0.12^{+0.04}_{-0.04} 
\\ 
 $0$        & $f$      & -1.4^{+2.6}_{-2.7}      &  5.38   
                       & -3.6^{+1.7}_{-1.8}      & 13.69   
                       &  0.03                   &  0.00 
\\
 $0$        & $fx$     & -1.9^{+3.7}_{-3.8}      &  5.40   
                       & -5.7\pm2.7              & 14.39   
                       &  0.05                   &  0.01 
\\
 $0$        & $f(1-x)$ & -4.9^{+8.6}_{-8.3}      &  5.38   
                       & -9.2^{+4.7}_{-7.9}      & 13.69   
                       &  0.0                    & -0.01  
\\
 $f(1-x)$   & $f$      & -3.5^{+2.5}_{-2.2}      &  3.35   
                       & -2.2^{+0.7}_{-0.8}      &  6.54   
                       &  0.01                   & -0.29\pm0.07  
\\
\end{tabular}
\end{ruledtabular}
\end{table*}

From the quadratic dependence of the branching ratio on form factors
one can derive the maximal reduction of the
$\pi\to e\nu\gamma$ branching ratio which may be caused by $f_T$ or $f_S$   
\begin{eqnarray}
   \Delta\mathrm{BR}^S_{\max} & = & -0.19~~~(f_S = -3.81)  \\
   \Delta\mathrm{BR}^T_{\max} & = & -0.82~~~(f_T = -1.95)
\end{eqnarray}
One can see that the amplitude $\cal S$ with constant form factor $f_S$
can not explain the ISTRA result. The possible
dependencies of $f_S$ on $x$, {\em e.g.} $f_S=f_S^{(0)}x$ or
$f_S=f_S^{(0)}(1-x)$, does not change this conclusion. Therefore, the low
value of the $\pi\to e\nu\gamma$ branching ratio, measured by ISTRA
experiment implies the presence of the tensor amplitude $\cal
T$ with a negative form factor.

To evaluate a possible $x$-dependence of the form factor $f_T(x)$ we can
simulate the available experimental data. For experiments
\cite{cern,berkeley,lampf,psi} a combined $\chi^2$ has been constructed as a
function of the ``true'' ratio of the axial to vector form factors
$\gamma$ and a 
hypothetical form factor $f_{T,S}(x)$. Minimization of the $\chi^2$ allows one
to evaluate the significance of the $f_{T,S}$ for improving the
consistency of the experimental data. The details of the analysis are
described in Ref. \cite{confirm}. 

To simulate the ISTRA results, the assumed $IB$
and $SD^\pm$ event distribution was modified by the hypothetical $\cal T$ (or
$\cal S$) contribution and  then it was fit
by a linear combination of $IB$, $SD^+$, and $SD^-$. The branching ratios 
of $IB$ and $SD^-$ were then compared with the ISTRA
values $BR_{IB}=1.62\pm0.20$ and $BR_{SD^-}=-0.58\pm0.20$ (in
units of $10^{-7}$) using the $\chi^2$ function
$\chi^2=(\delta_{IB}^2+\delta_{SD^-}^2+2\rho\delta_{IB}\delta_{SD^-})/(1-\rho^2)$,
where $\delta_{IB}=(BR_{IB}-1.62)/0.20$,
$\delta_{SD^-}=(BR_{SD^-}+0.58)/0.20$ and $\rho=0.67$ is the
estimate of the  correlation factor. The dependence of $\chi^2$
on the studied form factor, allows us to simulate the result of
the determination of this form factor with ISTRA data.

The form factors obtained in the ISTRA fit simulation were used to predict
the results of the PIBETA fit in the kinematic region
$E_e>20~\mathrm{MeV}$, $E_\gamma>55.6~\mathrm{MeV}$.

In the simulation of the ISTRA and PIBETA fits, the
detection efficiency was assumed to be constant in the phase
space regions of interest. The stability of the results against the possible
efficiency dependence on $x$ and $\lambda$ was specifically checked.
$\gamma=0.5$ was assumed in both simulations.
 
Several dependencies of $f_T$ and $f_S$ on $x$ were tested.
The correlated contribution of the form factors $f_T=f(1-x)$ and
$f_S=f$ may be associated with amplitude
$(k^\mu\epsilon^\nu-k^\nu\epsilon^\mu)q^\mu q^\rho L^{\rho\nu}$.
The results are summarized in Table \ref{tabl}.

One can see that form factor $f_S$ can not improve the consistency of the
experiments \cite{cern,berkeley,lampf,psi}, does not fit well the
ISTRA result, and does not effect the PIBETA fit. In other words,
there is no evidence of the structure dependent amplitude $\cal S$ in
the available experimental data.

The tensor form factors $f_T=f$ and $f_T=fx$ almost
perfectly simulate the ISTRA results. However they strongly disagree 
with the preliminary PIBETA fit.

A form factor $f_T=f(1-x)$ is not in good consistency with the
ISTRA data. Its value found in the ISTRA simulation fairly
well reproduces the total deficit of the $\pi\to e\nu\gamma$ decays.
However, the distribution of the simulated missing events is better 
approximated
by the $IB(x,\lambda)$ than by the $SD^-$ distribution. 
Taking into account the correlation between the 
$IB(x,\lambda)$ and $SD^-(x,\lambda)$ distributions, the low
statistics of the ISTRA experiment (about 100 events), and 
the possible uncertainties in the simulation procedure,
we should not rule out form factor $f_T=f(1-x)$ from the
consideration. In addition, this form factor seems to be the only one
which does not strongly contradicts both the ISTRA and the preliminary
PIBETA results. 

We may also point out that the value of $\gamma$ which one can obtain in the 
PIBETA preliminary fit has only a weak dependence on the actual form factors
$f_T(x)$ or $f_S(x)$.

The value of the constant tensor form factor $F_T=-0.0115\pm0.0033$
($f_T=-1.7\pm0.5$) obtained in the discussed simulation of the ISTRA fit
is by factor of two larger in absolute value than the estimate
in Ref. \cite{tensor}. In Ref. \cite{tensor}, $F_T=-0.0056\pm0.0017$ was 
estimated by
comparing the measured $BR_{SD^-}$ with the calculated
interference between tensor and $IB$ amplitudes. In this paper we took
into account the terms quadratic in $F_T$ and required the part of
total contribution of $F_T$ which may be approximated by functions
$IB(x,\lambda)$ and $SD^-(x,\lambda)$ to equal the measured sum of $IB$
and $SD^-$ contribution. This promises to be a better approach.

\begin{figure}
\begin{center}
\mbox{\includegraphics[width=0.4\textwidth]{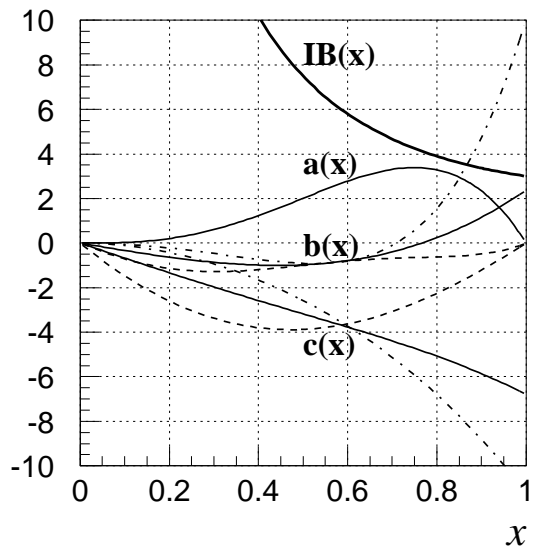}}
\end{center}
\caption{The dependence of the functions $a(x)$, $b(x)$, $c(x)$ on
$x$, assuming $f_V=3.78$, $f_A=0.5f_V$, and $f_S=0$. For the functions
$b(x)$ and $c(x)$, solid lines correspond to  $f_T=-1.7$, dashed
lines to  $f_T=-4.1(1-x)$, and dashed-dotted line to 
$f_T=-2.8x$.
The inner bremsstrahlung contribution $IB(x)$ is displayed
for comparison.}
\label{abc}
\end{figure}

It is needless to emphasize that the indirect analysis described above can not
substitute the real fit to the experimental data with the well determined
detector efficiencies and resolution.
One of the main goals of this paper is to stress that the analysis  of the
event distributions over $\lambda$ may provide model independent
determination of the form factors. As one can see from
Eq. (\ref{rate}), each contribution to the decay rate have the same
(known) $\lambda$-distribution independent of the value of $x$ and
the actual value of the form factors. 
Since $1-\lambda=(1-\lambda)^2 + \lambda(1-\lambda)$, only three
values may be independently determined for each $x$
\begin{eqnarray}
  \!\!\!\!a(x)\!&\!=\!&\!|f_V+f_A|^2\,x^3(1-x)                                \\
  \!\!\!\!b(x)\!&\!=\!&\!  2x\left[f_T+f_T^2x^2+(f_S+f_S^2x^2)(1-x)^2\right] 
\label{b(x)}   \\       
  \!\!\!\!c(x)\!&\!=\!&\!x\left[|f_V\!-\!f_A|^2x^2(1\!-\!x) +  4f_T+4f_S(1\!-\!x)^2\right]
\label{c(x)}  
\end{eqnarray}
from the analysis of the events distribution over $\lambda$,
\begin{equation}
    dN(x)/d\lambda~\propto~ a(x)\lambda^2+2b(x)\lambda(1-\lambda)+c(x)(1-\lambda)^2
\end{equation}
We assume that the inner bremsstrahlung contribution
$IB(x)(1-\lambda)/(3\lambda)$
with $IB(x)=3[(1-x)^2+1]/x$ \cite{IBcomment} is properly subtracted
from the experimental data.
Obviously, the nonzero value of $b(x)$ or negative value $c(x)$ is
evidence for the presence of the anomalous amplitudes $\cal T$
and/or $\cal S$. Also, $c(1)$ is directly related to the tensor form
factor $f_T^{(q^2=0)}=c(1)/4$ at $q^2=0$.
To illustrate the possible sensitivity of the method,
the functions of $a(x)$, $b(x)$, $c(x)$ are shown in Fig. \ref{abc}.
Functions $b(x)$ and $c(x)$ were calculated for three different tensor
form factors obtained from the simulation of ISTRA fit:
$f_T=-1.7$, $f_T=-2.8x$, and $f_T=-4.1(1-x)$. 
Fig. \ref{abc} shows these three cases are indistinguishable for
$x\approx0.6$. However, the measurement at $x>0.7$ allows one to make a
choice.

The measurement of $a(x)$ allows one to determine the  $|f_V(x)+f_A(x)|$ as
a function of $x$. Assuming that the vector form factor is known from the
CVC hypothesis and chiral perturbation theory calculations \cite{bijnens,geng}
one can extract, in principal,  $f_A(x)$, $f_T(x)$, and $f_S(x)$ from the 
measured values $a(x)$, $b(x)$, and $c(x)$.

In the above analysis of the $\pi\to e\nu\gamma$ decay rate, it was implicitly assumed 
that the form factors $f_T$ and $f_S$ are real and that there is no right
handed neutrino in the decay amplitude. Such additional
possibilities may be easily accounted for by the following modification of
the form factors in Eqs. (\ref{rate},\ref{b(x)},\ref{c(x)}):
\begin{eqnarray}
   f_{T,S}   & \to &  Re(f_{T,S})                     \\
   f_{T,S}^2 & \to &  |Re(f_{T,S})|^2 + |Im(f_{T,S})|^2 + |f_{T,S}^{(R)}|^2
\end{eqnarray}
Here, $f_{T,S}^{(R)}$ are the form factors of the amplitudes with right-handed neutrino. Alternatively, $f_{T,S}$ and $f_{T,S}^2$ may be considered as independent variables constrained only by the inequality
$f_{T,S}^2\ge f_{T,S}\times f_{T,S}$. For the tensor form factor $f_T^{(q^2=0)}$ at $q^2=0$, this inequality may be tested by the measurement of  
$b(1)$ and $c(1)$.
With such an extension of the decay amplitude we can no longer
determine all form factors in a model independent way. However, the
functions $a(x)$, $b(x)$, and $c(x)$ still contain the complete
experimental information which may be used in further theoretical
analysis.
In other words, the functions $a(x)/x^3(1-x)$, $b(x)/2x$, and $c(x)/x$
provide the model independent representation of the results of the $\pi\to
e\nu\gamma$ experiment. Neglecting radiative corrections and corrections due to the charged lepton mass, $m_e$, this statement is valid in the general case 
and may be violated only by the hypothetical dependence of the form factors on 
on the $e$-$\gamma$ opening angle.

The discussion of the possible sources of the anomalous amplitudes
$\cal S$ and $\cal T$ are beyond the scope this paper. Here, we will
only mention that numerous attempts to explain the ISTRA
result in the framework of the Standard Model, {\em e.g.} due to the radiative
corrections to the $\pi\to e\nu\gamma$ decay \cite{nikitin}, were
unsuccessful.

In this paper the most general $\pi \to e\nu\gamma$ amplitude was
considered. The model independent formulation of the results of the 
high statistics
and high resolution experiments is suggested. If the final PIBETA result
will validate the deficit of the $\pi\to e\nu\gamma$ decays observed
by the 
ISTRA experiment, it will be a solid argument in favor of presence of
the tensor component in the $\pi\to e\nu\gamma$ structure dependent
amplitude. Comparison of the available experimental data with the
preliminary PIBETA results indicate a possible strong dependence  of
tensor form factor on $q^2$.
The large  statistics accumulated by the PIBETA experiment allows one
to expect that the problem of the $\pi\to e\nu\gamma$ decay
will be resolved soon.

\begin{acknowledgments}
Author would like to thank D.~Lazarus for reading the manuscript and useful comments.
\end{acknowledgments}

\bibliography{pienug}

\begin{thebibliography}{21}
\bibitem[{IBc()}]{IBcomment}
\bibinfo{note}{Factor 3 is included to the definition of $IB(x)$ to have
  approximately the same normalization as for $a(x)$, $b(x)$, $c(x)$.}
\end{thebibliography}
\end{document}